\documentclass{article}

\usepackage{amsfonts}
\usepackage{amsmath}
\usepackage{amsthm}
\usepackage{bbm}
\usepackage{amssymb}
\usepackage{graphicx}
\usepackage{textcomp}
\usepackage[matrix,arrow,curve]{xy}
\usepackage{color}

\newtheorem{theorem}{Theorem}[section]

\newtheorem{example}[theorem]{Example}

\newtheorem{definition}[theorem]{Definition}

\def\be{\begin{eqnarray}}
\def\ee{\end{eqnarray}}

\newcommand{\ZG}{\mathbb{Z}}

\newcommand{\br}[1]{\left(#1\right)}

\newcommand{\eq}[1]{\begin{equation} #1 \end{equation}}

\newcommand{\eqm}[1]{\begin{multline} #1 \end{multline}}

\newcommand{\eql}[2]{\begin{equation} \label{e:#1} #2 \end{equation}}

\newcommand{\eqs}[1]{\begin{align} #1 \end{align}}

\newcommand{\re}[1]{(\ref{e:#1})}

\newcommand{\ths}{\vartheta}

\newcommand{\ts}[1]{\ifcase#1 \ths_{\br{D_8\oplus D_8}^+}\or \ths_{\ZG\oplus
A_{15}^+}\or\ths_{\ZG^2\oplus \br{E_7\oplus E_7}^+}\or\ths_{\ZG^4\oplus
D_{12}^+}\or \ths_{\ZG^8\oplus E_{8}}\or \ths_{\ZG^{16}} \or \ths_{E_8 \oplus
E_8} \or \ths_{D_{16}^+} \fi}
\DeclareMathOperator{\tr}{tr}
\DeclareMathOperator{\Li}{Li}

\DeclareMathOperator{\Ob}{O}

\newcommand{\ed}{\mathbbm{1}}

\newcommand{\NNN}[1]{ #1 }
\newcommand{\RRR}[1]{ #1 }

\hoffset= - 1.2in         

\title{{\bf Explicit computation of Drinfeld associator in the case of the fundamental representation of gl(N)} \vspace{.2cm}}
\author{{\bf P.Dunin-Barkowski}\thanks{{\small
{\it ITEP, Moscow, Russia and Korteweg-de Vries Institute for Mathematics, University of Amsterdam, The Netherlands}};
barkovs@itep.ru}, {\bf A.Sleptsov}\thanks{{\small
{\it ITEP, Moscow, Russia}};
sleptsov@itep.ru}, {\bf A.Smirnov}\thanks{{\small
{\it ITEP, Moscow, Russia and Columbia University, New York, USA}}; asmirnov@itep.ru}
\date{ }}

\begin{document}
\maketitle
\vspace{-5.0cm}
\begin{center}
\hfill ITEP/TH-64/11\\
\end{center}

\vspace{3.5cm}






\bigskip

\centerline{ABSTRACT}

\bigskip

{\footnotesize
We solve the regularized Knizhnik-Zamolodchikov equation and find an explicit expression for the Drinfeld associator. We restrict to the case of the fundamental representation of $gl(N)$. Several tests of the results are presented. It can be explicitly seen that components of this solution for the associator coincide with certain components of WZW conformal block for primary fields. We introduce the symmetrized version of the Drinfeld associator by dropping the odd terms. The symmetrized associator gives the same knot invariants, but has a simpler structure and is fully characterized by one symmetric function which we call the Drinfeld prepotential.
}

\bigskip

\section{Introduction}
It is well known that the polynomial invariants of knots can be evaluated with the help of Chern-Simons theory \cite{WCS}. We recall some essential facts and constructions.

The Chern-Simons action is given by $S_{CS}=\int tr_R\left(A\wedge dA+\frac{2}{3}A\wedge A\wedge A \right)$. Chern-Simons theory is a three-dimensional topological quantum field theory with gauge group $G$; traces are taken in a fixed representation $R$ of this group. Throughout this paper we let $G=GL(N)$.  One can consider the gauge invariant Wilson loop operators, which are defined as follows:
\be
\label{vev}
\langle \, W(K)\,\rangle= \frac{1}{Z} \int DA \, \exp\Big(-\frac{i}{\hbar} S_{CS}(A)\Big)\, W(C,A)
\ee
where
$$
Z=\int DA\,\exp\Big(-\frac{i}{\hbar} S_{CS}(A) \Big), \ \ \  W(C,A)=\tr_R {\rm Pexp} \oint\limits_{C} A
$$
The vacuum expectation value (\ref{vev}) does not depend on the realization $C$ of the knot in $\mathbb{R}^3$, but only on the equivalence class of the knot $K$ therefore,  $\langle\, W(K)\,\rangle$ defines a knot invariant. Then in order to evaluate the integral (\ref{vev}) one fixes some particular gauge. There are two relevant gauges: the holomorphic gauge \cite{La6, DBSS} and the temporal gauge \cite{Sm1, Sm2}.

If the temporal gauge is chosen then integral (\ref{vev}) gives a polynomial invariant, namely it is proportional to the HOMFLY polynomial which specializes to the Jones polynomial if one fixes a dimension $N=2$ gauge group. In the holomorphic gauge case, the vacuum expectation value (\ref{vev}) is equal to the celebrated Kontsevich integral. Due to the gauge invariance of Chern-Simons theory these answers should coincide, i.e. the Kontsevich integral computed for a particular knot (or link) is equal to the corresponding HOMFLY polynomial  of this knot (or link). However the Kontsevich integral is given by a complicated infinite power series (\cite{ChD, DBSS}). It is impossible to sum  this series in general and thus it is difficult to check this explicitly.

 On the other hand the Kontsevich integral for knots and links can be calculated as a trace (over a representation $R$ of the gauge group $G$) of a certain product of two basic operators: an R-matrix and a Drinfeld associator $\Phi$ \cite{ChD, ChDuBook, DBSS}. The R-matrix has a very simple structure, while the associator looks almost as complicated as the Kontsevich integral itself. It is given by an infinite power series in two non-commuting variables $A$ and $B$ with multiple zeta values as coefficients. In order to compute the Kontsevich integral in this way one has to sum the series for the associator. This paper addresses the problem of summing the associator series for the simplest case of the fundamental representation (see also \cite{Le, Le2}). The key is the Knizhnik-Zamolodchikov equation, which is an additional constraint satisfied by the correlation function of the conformal field theory associated with an affine Lie algebra at a fixed level (see \cite{Drin}). The solutions to this equation are
the correlation functions of primary fields (see \cite{diFran}). However, in practice they are rather difficult to solve directly, except for the four-point functions, which are directly related to associators \cite{ChD}. The only difference between associators and these four-point functions is that the associator is a solution of a regularized version of the equation. In addition, there are actually three slightly different versions of associators and their corresponding equations.

The problem of explicitly evaluating the Kontsevich integral may seem to be just a technical problem. However we emphasize that the search of explicit expressions for the associator $\Phi$ is very important and is a conceptual problem. The point is that the associator plays a key role in many areas of mathematics and theoretical physics such as Hopf algebras, conformal field theory, quantum groups, braid theory and many others.

In this paper we present an explicit expression for all forms of the associator in the case of the fundamental representation of $gl(N)$. The coefficients of one of these forms turn out to  coincide with certain components of the WZW conformal block for primary fields. We apply the obtained results to a few simple knots to check that the expression for the Wilson loop in the two gauges mentioned above actually coincide.

The odd part of the associator is represented by the values of multiple zeta functions at some odd points, and therefore does not enter the expression for knot invariants. We show that dropping the odd terms leads to a symmetrized version of the Drinfeld associator whose structure is simpler: it is fully characterized by one function even in Planck constant. We call it the Drinfeld prepotential.

Our paper is organized as follows. In Section \ref{sec:KZA} we introduce the Knizhnik-Zamolodchikov equation and its regularized form, and then we list three different forms of this equation needed for the knot theory. In Section \ref{sec:Sol} we provide explicit solutions to all forms of the equation from the previous section, for the case of the fundamental representation of $gl(N)$. In Section \ref{sec:Tests} we apply the obtained results to several particular examples of knots and find that they lead to invariants coinciding with the HOMFLY polynomials. In the final section, we introduce the symmetrized version of the associator which does not contain odd zeta function values.

\section{Regularized Knizhnik-Zamolodchikov equation for associator}
\label{sec:KZA}
Associators appearing in combinatorial construction of Kontsevich integral coefficients are related to the so-called Knizhnik-Zamolodchikov equation. This is an equation coming from conformal field theory \cite{diFran}. Consider the four-point function in WZW model:
$$
G(z)=\langle\, g(0) g(1) g(z) g(\infty)\,  \rangle_{WZW}
$$
Knizhnik-Zamolodchikov equation is the Ward identity for this four-point function:
\be
\label{KZ}
\dfrac{d G(z)}{d z} =\hbar\,\Big( \dfrac{A}{z}+\dfrac{B}{1-z} \Big)\,G(z)
\ee
where $A$ and $B$ have the form (we assume summation over repetitive indices):
$$
A=T^a\otimes \left(T^a\right)^* \otimes \ed,\ \ \ B=\ed\otimes T^a\otimes \left(T^a\right)^*
$$
\NNN{ $G(z)$ is a rank-six tensor of the same type as $A$ and $B$. Also note that everywhere where we write a product of two such rank-six tensors, we assume the following:
\eql{prodcon}{
(AB)^{\alpha\beta\gamma}_{\delta\epsilon\chi} := A^{\alpha\beta\gamma}_{\mu\nu\rho}B^{\mu\nu\rho}_{\delta\epsilon\chi}
}
}
The perturbative approach for solving of the equation (\ref{KZ}) is discussed in Appendix \ref{a:pert}. The solution in the first two orders of perturbative expansion is also given. \NNN{Solutions of $sl_N$ Knizhnik-Zamolodchikov equations at level zero are given in \cite{AN}. However we are interested in general situation and in regularized Knizhnik-Zamolodchikov equation. Moreover we need explicit expression for assocaitor to apply the result to the knot theory.}

This solution $G(z)$ diverges at points $0$ and $1$, so we
make the following transformation to eliminate the divergences:
\eq{
G(z) = \br{1-z}^{\NNN{-}\hbar B}\Phi(z) z^{\hbar A}.
}
The regularized equation has then the form
\eql{phi}{
\begin{array}{|c|}
\hline\\
\dfrac{d\Phi}{dz} = \dfrac{\hbar}{z}\,\br{1-z}^{\hbar B} A\, \br{1-z}^{\hbar
B}\, \Phi - \dfrac{\hbar}{z}\,\Phi A.\\ \\
\hline
\end{array}
}
The solution of this equation taken at point $z=1$ gives us the associator.
In Section \ref{sec:Sol} equation \re{phi} is solved in the case of fundamental representation of $gl(N)$.

\subsection{\NNN{T}ypes of associators}
Actually, to construct the coefficients of the Kontsevich integral combinatorially (see \cite{ChD, DBSS}), one needs the $R$-matrix and three types of associators. These three types of associators correspond to three  types of orientation of strands of a braid and lead to three slightly different forms of equation \re{phi}. They are listed in the following table:
\eq
{\label{tables}
\begin{array}{|c|c|c|c|}
\hline
\rm{Orientation} & \rm{KZ-equation} & \rm{Transformation} & \rm{Associator} \
\rm{equation} \\
\hline
 & & & \\
$\bf{a}) \ \Big\downarrow\Big\downarrow\Big\downarrow$ & \dfrac{d G(z)}{d z} =\hbar\,\Big(
\dfrac{A}{z}-\dfrac{B}{1-z} \Big)\,G(z) & G(z) = \br{1-z}^{\NNN{\hbar} B}\Phi(z) z^{\NNN{\hbar} A} &
\dfrac{d\Phi}{dz} = \dfrac{\hbar}{z}\,\br{1-z}^{-\hbar B} A\, \br{1-z}^{\hbar
B}\, \Phi - \dfrac{\hbar}{z}\,\Phi A \\
& & & \\
\hline
& & & \\
$\bf{b}) \ \Big\downarrow\Big\uparrow\Big\downarrow$ & \dfrac{d G(z)}{d z} =-\hbar\,\Big(
\dfrac{A}{z}-\dfrac{B}{1-z} \Big)\,G(z) & G(z) = \br{1-z}^{-\NNN{\hbar} B}\Phi(z) z^{-\NNN{\hbar} A} &
\dfrac{d\Phi}{dz} = \dfrac{-\hbar}{z}\,\br{1-z}^{\hbar B} A\, \br{1-z}^{-\hbar
B}\, \Phi + \dfrac{\hbar}{z}\,\Phi A \\
& & & \\
\hline
& & & \\
$\bf{c}) \ \Big\downarrow\Big\downarrow\Big\uparrow$ & \dfrac{d G(z)}{d z} =\hbar\,\Big(
\dfrac{A}{z}+\dfrac{B}{1-z} \Big)\,G(z) & G(z) = \br{1-z}^{-\NNN{\hbar} B}\Phi(z) z^{\NNN{\hbar} A} &
\dfrac{d\Phi}{dz} = \dfrac{\hbar}{z}\,\br{1-z}^{\hbar B} A\, \br{1-z}^{-\hbar
B}\, \Phi - \dfrac{\hbar}{z}\,\Phi A \\
& & & \\
\hline
\end{array}
}

\section{Solutions}
\label{sec:Sol}
In this section we solve the corresponding equation for each of three cases (\ref{tables}). We restrict our considerations to the case of fundamental representation of $gl(N)$ algebra only.

\subsection{Type a}
Let us consider type \textbf{a} from table \ref{tables}. In this case $A$ and $B$ take the following form
\NNN{
\eq{
\label{gens}
A=T^a\otimes \left(T^a\right)^* \otimes \ed,\ \ \ B=\ed\otimes T^a\otimes \left(T^a\right)^*,
}
where, since we are considering the fundamental representation of $gl(N)$,
\eq{
\label{genfund}
T^a=g_{ij}, \ \left(T^a\right)^*=g_{ji},
}
where index $a$ corresponds to a pair of indices $i$ and $j$ each of which goes from 1 to $N$, and $g_{ij}$ is a matrix with components equal to $0$ except the $(i,j)$-th one, which is equal to $1$.
}
Then we get the following relations for $A$ and $B$:
\eq{
A^2=B^2=\br{AB}^3=\br{BA}^3=1. \label{rel}
}
\NNN{In the paper \cite{Le} it is proven that associator $\Phi$ is given by the infinite series in $A$ and $B$ (\ref{ass}). However there are relations (\ref{rel}) on $A$ and $B$, therefore associator can be written as finite polynomial:
\eq{ \label{phipol}
\Phi = \Phi_1 + \Phi_2 A + \Phi_3 B + \Phi_4 AB +\Phi_5 BA + \Phi_6 ABA,
}
where $\Phi_i$ are some coefficients and do not depend on $A$ and $B$. Now let us substitute (\ref{phipol}) in equation \re{phi}, then it turns into a system of differential equations, which can be represented in the matrix form:
\eql{phieq}{
\dfrac{d\vec\Phi}{dz} = \dfrac{\hbar}{z}\, M(z)\vec\Phi,
}
where
\eq{
\vec\Phi= \br{\begin{array}{c}
\Phi_1 \\
\Phi_2 \\
\Phi_3 \\
\Phi_4 \\
\Phi_5 \\
\Phi_6
 \end{array}}.
}
}

Matrix $M$ has the form:
\eq{
M = \br{\begin{array}{cccccc}
0 & s^2 & 0 & -cs & cs & -s^2 \\
s^2 & 0 & cs & -s^2 & 0 & -cs \\
0 & -cs & 0 & c^2 & -c^2 & cs \\
cs & -s^2 & c^2 & 0 & -cs & -1 \\
-cs & 0 & -c^2 & cs & 0 & c^2 \\
-s^2 & cs & -cs & -1 & c^2 & 0 \\
 \end{array}},
}
where
\eqs{
c := \cosh\br{\hbar \log\br{1-z}}=
\dfrac{1}{2}\br{\br{1-z}^{\hbar}+\br{1-z}^{-\hbar}},\\
s := \sinh\br{\hbar \log\br{1-z}}=
\dfrac{1}{2}\br{\br{1-z}^{\hbar}-\br{1-z}^{-\hbar}}.
}
In order to solve equation \re{phieq} one can change the basis in the following way:
\eql{phitran}{
\widetilde{\vec\Phi} = S \vec\Phi,
}
\eq{
S = \left(
\begin{array}{cccccc}
 1 & 0 & 0 & 1 & 1 & 0 \\
 0 & 1 & 1 & 0 & 0 & 1 \\
 -\sqrt{3} & -\sqrt{3} & \sqrt{3} & 0 & \sqrt{3} & 0 \\
 -1 & -1 & -1 & 2 & -1 & 2 \\
 \sqrt{3} & -\sqrt{3} & \sqrt{3} & 0 & -\sqrt{3} & 0 \\
 -1 & 1 & 1 & 2 & -1 & -2
\end{array}
\right)
}
Then matrix $M$ takes the form
\eq{
\widetilde{M}=SMS^{-1}=
\left(
\begin{array}{cccccc}
 0 & 0 & 0 & 0 & 0 & 0 \\
 0 & 0 & 0 & 0 & 0 & 0 \\
 0 & 0 & -\frac{1}{2} & \frac{\sqrt{3}}{2} \br{1-z}^{2\hbar} & 0 & 0 \\
 0 & 0 & \frac{\sqrt{3}}{2} \br{1-z}^{-2\hbar} & -\frac{3}{2} & 0 & 0 \\
 0 & 0 & 0 & 0 & \frac{1}{2} & \frac{\sqrt{3}}{2} \br{1-z}^{-2\hbar} \\
 0 & 0 & 0 & 0 & \frac{\sqrt{3}}{2} \br{1-z}^{2\hbar} & \frac{3}{2}
\end{array}
\right)
}
Hence, matrix equation \re{phieq} turns into the following system of differential equations:
\eqs{ \label{eqsf}
\left\{
\begin{array}{l}
z\dfrac{d\widetilde{\Phi_3}}{dz}=-\frac{1}{2}\RRR{\hbar}\widetilde{\Phi_3}+\frac{\sqrt{3}}{2} \RRR{\hbar}\br{1-z}^{2\hbar}\widetilde{\Phi_4} ,\\[4mm]
z\dfrac{d\widetilde{\Phi_4}}{dz}=-\frac{3}{2}\RRR{\hbar}\widetilde{\Phi_4}+\frac{\sqrt{3}}{2}\RRR{\hbar} \br{1-z}^{-2\hbar}\widetilde{\Phi_3}, \\[4mm]
z\dfrac{d\widetilde{\Phi_5}}{dz}=\frac{1}{2}\RRR{\hbar}\widetilde{\Phi_5}+\frac{\sqrt{3}}{2} \RRR{\hbar}\br{1-z}^{-2\hbar}\widetilde{\Phi_6} ,\\[4mm]
z\dfrac{d\widetilde{\Phi_6}}{dz}=\frac{3}{2}\RRR{\hbar}\widetilde{\Phi_6}+\frac{\sqrt{3}}{2}\RRR{\hbar} \br{1-z}^{2\hbar}\widetilde{\Phi_5} \\[4mm]
\end{array}
\right.
}
with the following initial conditions $\br{\Phi(0)=1}$:\NNN{
\eqs{ \label{conds}
\left\{
\begin{array}{l}
\widetilde{\Phi_3}(0)=-\sqrt{3}, \\
\widetilde{\Phi_4}(0)=-1, \\
\widetilde{\Phi_5}(0)=\sqrt{3}, \\
\widetilde{\Phi_6}(0)=-1
\end{array}
\right.
}
From (\ref{eqsf}) and (\ref{conds}) one can obtain conditions for derivatives of $\dot{\widetilde{\Phi_i}}(0)$ using $\dot{\widetilde{\Phi}}(0)=\RRR{\hbar}\widetilde{M}(0)\dot{\widetilde{\Phi}}(0)+\RRR{\hbar}\dot{\widetilde{M}}(0)\widetilde{\Phi}(0)$:
\RRR{
\eqs{ \label{conds2}
\left\{
\begin{array}{l}
\dot{\widetilde{\Phi_3}}(0)=\dfrac{\sqrt{3} \hbar^2}{1+2 \hbar}, \\
\dot{\widetilde{\Phi_4}}(0)=-\dfrac{3 \hbar^2}{1+2 \hbar}, \\
\dot{\widetilde{\Phi_5}}(0)=\dfrac{\sqrt{3} \hbar^2}{-1+2 \hbar}, \\
\dot{\widetilde{\Phi_6}}(0)=\dfrac{3 \hbar^2}{-1+2 \hbar}
\end{array}
\right.
}
}
Thus systems of equations (\ref{eqsf}-\ref{conds2}) can be reduced to a system of hypergeometric
equations:
\eq{
\left\{
\begin{array}{l}
z\br{1-z}\ddot{\widetilde{\Phi_3}} + \br{1+2\hbar-z}\dot{\widetilde{\Phi_3}} + \hbar^2\widetilde{\Phi_3} = 0 ,\\[4mm]
z\br{1-z}\ddot{\widetilde{\Phi_4}} + \br{1+2\hbar-\br{1+4\hbar}z}\dot{\widetilde{\Phi_4}} - 3\hbar^2\widetilde{\Phi_4} = 0 \\[4mm]
z\br{1-z}\ddot{\widetilde{\Phi_5}} + \br{1-2\hbar-z}\dot{\widetilde{\Phi_5}} + \hbar^2\widetilde{\Phi_5} = 0 ,\\[4mm]
z\br{1-z}\ddot{\widetilde{\Phi_6}} + \br{1-2\hbar-\br{1-4\hbar}z}\dot{\widetilde{\Phi_6}} - 3\hbar^2\widetilde{\Phi_6} = 0, \\[4mm]
\end{array}
\right.
}
We get a system of differntial equations of the second order. Solutions of this system considered with initial conditions (\ref{conds}, \ref{conds2}) are equivalent to solutions of (\ref{eqsf}). Solutions are given by the following hypergeometric functions:
\eq{ \label{ans1}
\left\{
\begin{array}{l}
\widetilde{\Phi_1} = 1 ,\\[4mm]
\widetilde{\Phi_2} = 0 ,\\[4mm]
\widetilde{\Phi_3} = -\sqrt{3} \ \RRR{_2}F_1\br{\hbar,-\hbar,1+2\hbar;z} ,\\[4mm]
\widetilde{\Phi_4} = -            \ \RRR{_2}F_1\br{\hbar,3\hbar,1+2\hbar;z} ,\\[4mm]
\widetilde{\Phi_5} = \sqrt{3}   \ \RRR{_2}F_1\br{\hbar,-\hbar,1-2\hbar;z}, \\[4mm]
\widetilde{\Phi_6} = -            \ \RRR{_2}F_1\br{-\hbar,-3\hbar,1-2\hbar;z} \\[4mm]
\end{array}
\right.
}
\RRR{We have checked with the help of a computer algebra system that solutions (\ref{ans1}) indeed obey original equations (\ref{eqsf}) with initial conditions (\ref{conds}).}

Thus we see that coefficients of associator of this type are given by linear combinations of values of hypergeometric functions taken at the point $z=1$, which can be rewritten in terms of Gamma function and trigonometric functions. Note that the answer in this case does not depend on order $N$ of algebra $gl(N)$:
\eq{ \label{ans11}
\left\{
\begin{array}{l}
\widetilde{\Phi_1}(1) = 1 ,\\[4mm]
\widetilde{\Phi_2}(1) = 0 ,\\[4mm]
\widetilde{\Phi_3}(1) = -\sqrt{3} \ \RRR{_2}F_1\br{\hbar,-\hbar,1+2\hbar;1} ,\\[4mm]
\widetilde{\Phi_4}(1) = -            \ \RRR{_2}F_1\br{\hbar,3\hbar,1+2\hbar;1} ,\\[4mm]
\widetilde{\Phi_5}(1) = \sqrt{3}   \ \RRR{_2}F_1\br{\hbar,-\hbar,1-2\hbar;1}, \\[4mm]
\widetilde{\Phi_6}(1) = -            \ \RRR{_2}F_1\br{-\hbar,-3\hbar,1-2\hbar;1} \\[4mm]
\end{array}
\right.
\Longleftrightarrow \ \
\left\{
\begin{array}{l}
\widetilde{\Phi_1}(1) = 1 ,\\[4mm]
\widetilde{\Phi_2}(1) = 0 ,\\[4mm]
\widetilde{\Phi_3}(1) = -\sqrt{3} \ \dfrac{\Gamma(1+2\hbar)^2}{\Gamma(1+\hbar)\Gamma(1+3\hbar)} ,\\[4mm]
\widetilde{\Phi_4}(1) = -              \ \dfrac{1}{{\rm cos}(\pi\hbar)} ,\\[4mm]
\widetilde{\Phi_5}(1) = \sqrt{3}   \ \dfrac{\Gamma(1-2\hbar)^2}{\Gamma(1-\hbar)\Gamma(1-3\hbar)}, \\[4mm]
\widetilde{\Phi_6}(1) = -              \ \dfrac{1}{{\rm cos}(\pi\hbar)}.\\[4mm]
\end{array}
\right.
}
Thus we obtained explicit expressions for $\widetilde{\vec \Phi}$. To get expression for associator one needs to take $\vec \Phi = S^{-1}\widetilde{\vec \Phi}$ and then use (\ref{phipol}).
}

\subsection{Type b}
Now we briefly present results for type \textbf{b} from table \ref{tables}. In this case we have:
\eql{genb}{
A=g_{ij}\otimes g_{ij}\otimes \ed, \ B=\ed\otimes g_{ij}\otimes g_{ij}
}
Relations for $A$ and $B$ are the following ones
\begin{equation}
\label{tb}
 \left\{
\begin{array}{l}
A^2 = NA ,\\[4mm]
B^2 =  NB, \\[4mm]
BAB = B ,\\[4mm]
ABA = A \\[4mm]
\end{array}
\right.
\end{equation}
Then the associator takes the form
\eql{dud1}{
\Phi = \Phi_1 + \Phi_2 A + \Phi_3 B + \Phi_4 AB +\Phi_5 BA.
}
By analogy with the previous case we write
\eql{dud2}{
\widetilde{\vec\Phi} = S \vec\Phi,
}
for
\eql{Smatrixb}{
S = \left(
\begin{array}{ccccc}
 1 & 0 & 0 & 0 & 0  \\
 1 & N & 0 & 1 & 0  \\
 \frac{1}{N^2-1} & 0 & 0 & -1 & 0  \\
 1 & 0 & N & 1 & 0  \\
 \frac{1}{N^2} & \frac{1}{N} & \frac{1}{N} & \frac{1}{N^2} & 1
\end{array}
\right)
}
Then we have
\eql{hyperb}{
\NNN{\left\{
\begin{array}{l}
\widetilde{\Phi_1} = 1 ,\\[4mm]
\widetilde{\Phi_2} =  \ \RRR{_2}F_1\br{\hbar,-\hbar,1-N\hbar;z}, \\[4mm]
\widetilde{\Phi_3} = \dfrac{1}{N^2-1} \ \RRR{_2}F_1\br{(N-1)\hbar,(N+1)\hbar,1+N\hbar;z} ,\\[4mm]
\widetilde{\Phi_4} =  \ \RRR{_2}F_1\br{\hbar,-\hbar,1+N\hbar;z} ,\\[4mm]
\widetilde{\Phi_5} = \dfrac{1}{N^2} \ \RRR{_2}F_1\br{-(N-1)\hbar,-(N+1)\hbar,1-N\hbar;z} \\[4mm]
\end{array}
\right.
\Rightarrow} \ \
\left\{
\begin{array}{l}
\widetilde{\Phi_1}(1) = 1 ,\\[4mm]
\widetilde{\Phi_2}(1) =  \ \dfrac{\Gamma(1-N\hbar)^2}{\Gamma(1-N\hbar-\hbar)\Gamma(1-N\hbar+\hbar)}, \\[4mm]
\widetilde{\Phi_3}(1) = \dfrac{1}{N^2-1} \ \dfrac{N{\rm sin}(\pi\hbar)}{{\rm sin}(\pi N\hbar)} ,\\[4mm]
\widetilde{\Phi_4}(1) =  \ \dfrac{\Gamma(1+N\hbar)^2}{\Gamma(1+N\hbar-\hbar)\Gamma(1+N\hbar+\hbar)} ,\\[4mm]
\widetilde{\Phi_5}(1) = \dfrac{1}{N^2} \ \dfrac{N{\rm sin}(\pi\hbar)}{{\rm sin}(\pi N\hbar)}. \\[4mm]
\end{array}
\right.
}

Here we also see that coefficients of associator of this type are given by linear combinations of values of hypergeometric functions taken at the point $z=1$.

Note that the coefficients of the associator in this case coincide with certain components of the WZW conformal block for primary fields in the fundamental representation. Namely, we have $F_1^{(-)}=\widetilde{\Phi_2}, \ F_1^{(+)}=\widetilde{\Phi_3}$ up to singular factors and substitution $\hbar = \dfrac{1}{k}$, where $F_1^{(-)}$ and $F_1^{(+)}$ are the components of the WZW conformal block for primary fields taken from Section 15.3.2 of \cite{diFran}. The other two coefficients $\widetilde{\Phi_4}$ and $\widetilde{\Phi_5}$ differ from the previous two just by substitution $N \rightarrow -N$.

Again, to get the expression for associator one needs to take $\vec \Phi = S^{-1}\widetilde{\vec \Phi}$ and then use \re{dud1}.

\subsection{Type c}
\begin{figure}[h!]
\begin{center}
\includegraphics[scale=0.2]{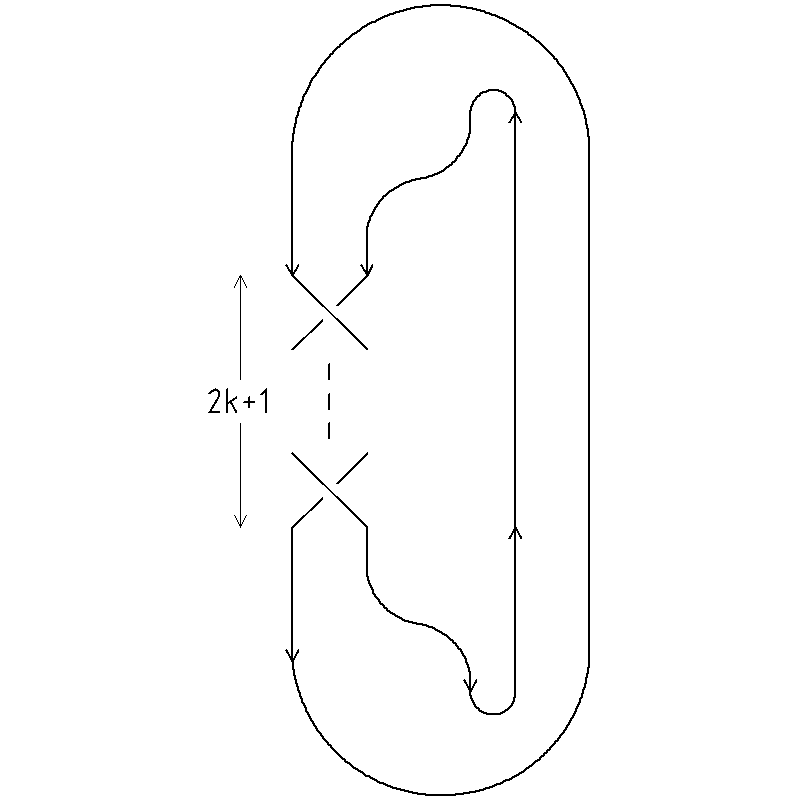}
\caption{(2k+1)-foil knot}
\label{tref}
\end{center}
\end{figure}

Associators of type \textbf{c} allow us to calculate the simplest knots like in Figure \ref{tref}.
In this case we have:
\eql{genc}{
A=g_{ij}\otimes g_{ji}\otimes \ed, \ B=\ed\otimes g_{ij}\otimes g_{ij}
}
Relations for $A$ and $B$ are the following ones:
\eql{relc}{ \left\{
\begin{array}{l}
A^2 = \NNN{1} ,\\[4mm]
B^2 =  NB, \\[4mm]
BAB = B ,\\[4mm]
\end{array}
\right.
}
Then the associator takes the form
\eql{ddu1}{
\Phi = \Phi_1 + \Phi_2 A + \Phi_3 B + \Phi_4 AB +\Phi_5 BA + \Phi_6 ABA.
}
By analogy with the previous case we write
\eql{ddu2}{
\widetilde{\vec\Phi} = S \vec\Phi,
}
for
\eql{Smatrixc}{
S = \left(
\begin{array}{cccccc}
 1 & 0 & 0 & 0 & 0 & 0  \\
 -N & 1 & 0 & 0 & 0 & 0  \\
 \frac{1}{2N} & \frac{1}{2N} & \frac{1}{2} & \frac{1}{2N} & \frac{1}{2} &
\frac{1}{2N}  \\
 \frac{1}{2(N+1)} & \frac{1}{2(N+1)} & 0 & \frac{1}{2} & 0 & \frac{1}{2}  \\
 -\frac{1}{N} & \frac{1}{N} & -1 & -\frac{1}{N} & 1 & \frac{1}{N} \\
\frac{1}{N-1} & -\frac{1}{N-1} & 0 & -1 & 0 & 1  \\
\end{array}
\right)
}
Then
\eql{hyperc}{
\NNN{\left\{
\begin{array}{l}
\widetilde{\Phi_1} = 1 ,\\[4mm]
\widetilde{\Phi_2} = -N, \\[4mm]
\widetilde{\Phi_3} = \dfrac{1}{2N} \ \RRR{_2}F_1\br{\hbar,-(N-1)\hbar,1+2\hbar;z}
,\\[4mm]
\widetilde{\Phi_4} =  \dfrac{1}{2(N+1)} \ \RRR{_2}F_1\br{\hbar,(N+1)\hbar,1+2\hbar;z}
,\\[4mm]
\widetilde{\Phi_5} = -\dfrac{1}{N} \ \RRR{_2}F_1\br{-\hbar,-(N+1)\hbar,1-2\hbar;z}
\\[4mm]
\widetilde{\Phi_6} = \dfrac{1}{N-1} \ \RRR{_2}F_1\br{-\hbar,(N-1)\hbar,1-2\hbar;z}
\\[4mm]
\end{array}
\right.
\Rightarrow} \ \
\left\{
\begin{array}{l}
\widetilde{\Phi_1}(1) = 1 ,\\[4mm]
\widetilde{\Phi_2}(1) = -N, \\[4mm]
\widetilde{\Phi_3}(1) = \dfrac{1}{2N} \ \dfrac{4^{\hbar}\Gamma(\hbar+\frac{1}{2})\Gamma(1+N\hbar)}{\sqrt{\pi}\Gamma(1+N\hbar+\hbar)},\\[4mm]
\widetilde{\Phi_4}(1) =  \dfrac{1}{2(N+1)} \ \dfrac{4^{\hbar}\Gamma(\hbar+\frac{1}{2})\Gamma(1-N\hbar)}{\sqrt{\pi}\Gamma(1-N\hbar+\hbar)},\\[4mm]
\widetilde{\Phi_5}(1) = -\dfrac{1}{N} \ \dfrac{4^{-\hbar}\Gamma(-\hbar+\frac{1}{2})\Gamma(1+N\hbar)}{\sqrt{\pi}\Gamma(1+N\hbar-\hbar)},\\[4mm]
\widetilde{\Phi_6}(1) = \dfrac{1}{N-1} \ \dfrac{4^{-\hbar}\Gamma(-\hbar+\frac{1}{2})\Gamma(1-N\hbar)}{\sqrt{\pi}\Gamma(1-N\hbar-\hbar)}.\\[4mm]
\end{array}
\right.
}

In this case as in previous ones coefficients of associator of this type are given by linear combinations of values of hypergeometric functions taken at the point $z=1$.

\section{\NNN{Comparison to known formulas for knots}}

\label{sec:Tests}
\NNN{In the present section we apply the formulas for the associator obtained in Section \ref{sec:Sol} to compute Kontsevich integral for a few knots. We do this following \cite{ChD,DBMMS}. The computations are done in the case of the fundamental representation of $gl(N)$. In this case Kontsevich integral is known to coincide with HOMFLY polynomial for the given knot, and we use this fact to check the validity of our formulas for the associator. Namely, we check that Kontsevich integral computed with the help of these formulas indeed coincides with known expressions for HOMFLY polynomial.}

\subsection{Hump}
\begin{figure}[h!]
\begin{center}
\includegraphics[scale=0.15]{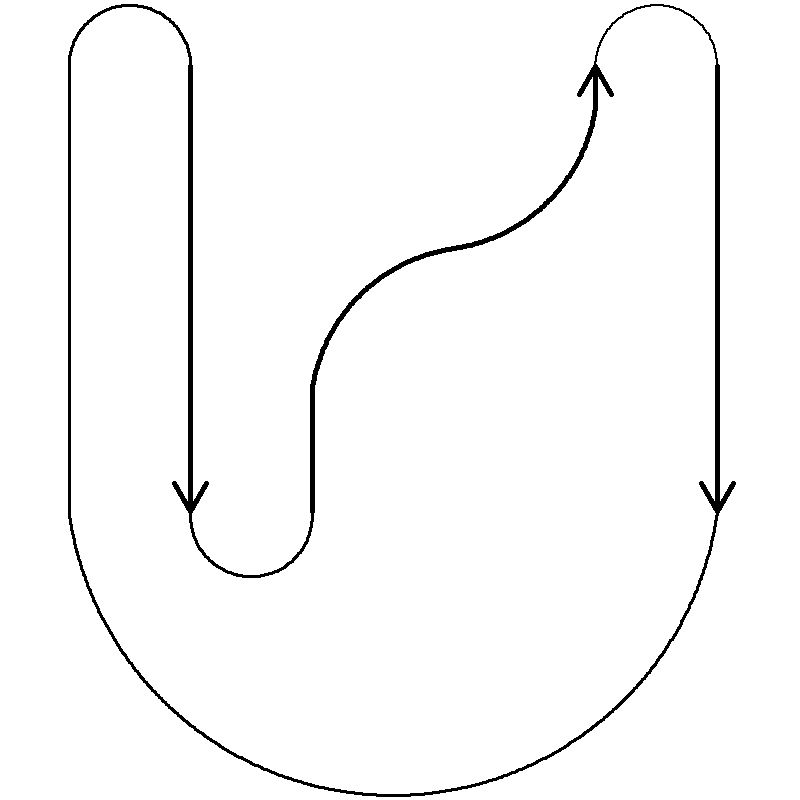}
\caption{Hump}
\label{hump}
\end{center}
\end{figure}

\NNN{We start with the \textit{hump}, which is a version of the \textit{unknot}}
(for the definition of the hump see \cite{ChD}).
\NNN{Recall that associator is a tensor with 6 indices. Denote the associtaor of type \textbf{b} as $\Phi_b$. Kontsevich integral of the hump corresponds, by definition, to the contraction of the indices of $\Phi_b$ corresponding to Figure \ref{hump}. We will refer to this contraction as the \textit{trace} of the assocaitor.}

\NNN{Writing this contraction in terms of the indices gives the following expression:
\eq{
KI_{\mbox{\tiny{Hump}}}= \tr_{\mbox{\tiny{Hump}}} \Phi_b := \mathop{\sum}_{\alpha,\beta,\gamma}(\Phi_b)^{\alpha\beta\beta}_{\gamma\gamma\alpha}
}
Now let us expend the associator as in formula \re{dud1}:
\eql{dud3}{
\Phi_b = \Phi_1 + \Phi_2 A + \Phi_3 B + \Phi_4 AB +\Phi_5 BA.
}
}

\NNN{
With the help of formula \re{genb} one can compute traces (in the above mentioned sense) of particular terms of \re{dud3} in the following way:}
\eqs{\label{trA}
\tr_{\NNN{\mbox{\tiny{Hump}}}} A = \dfrac{1}{N}\tr\br{g_{ij}g_{ji}} = \dfrac{1}{N}\tr\br{\delta_{jj}g_{ii}}
= N \\
\tr_{\NNN{\mbox{\tiny{Hump}}}} AB = \dfrac{1}{N}\tr\br{g_{ij}g_{nm}g_{ji}g_{mn}} = 1 \\
\tr_{\NNN{\mbox{\tiny{Hump}}}} BA = \dfrac{1}{N}\tr\br{g_{ij}g_{mn}g_{nm}g_{ji}} = N^2
\label{trBA}
}
\NNN{
Note that $\tr_{\mbox{\tiny{Hump}}}$ here stands for the above mentioned trace of a rank-six tensor, while $\tr$ stands for just the ordinary trace of a matrix. We also assume summation over repetitive indices.}

Now substituting formulas \re{hyperb}, \re{Smatrixb}, (\ref{trA})-(\ref{trBA})
to \re{dud2} we get the following answer:
\eql{qdim}{
KI_{Hump} = \ \RRR{_2}F_1\br{-(N-1)\hbar,-(N+1)\hbar,1-N\hbar;1} = N\cdot\dfrac{e^{\pi
i\hbar}-e^{-\pi i\hbar}}{e^{\pi iN\hbar}-e^{-\pi iN\hbar}} = \dfrac{N}{[N],}
}
where we introduce the following notation for a $q$-number $$[N] =
\dfrac{q^N-q^{-N}}{q-q^{-1}} , \ \ \ \ q=e^{\pi i\hbar}.$$

\NNN{This indeed coincides with the known result for the HOMFLY polynomial of the hump (which can be found e.g. in \cite{DBMMS}).}

\subsection{Series [2,2k+1]}
\NNN{Consider the so-called torus knots of type [2,2k+1], as shown in Figure \ref{tref}. Denote the associator of type \textbf{c} as $\Phi_c$. With the help of this associator one can compute Kontsevich integral for knots of this type \cite{ChD,DBMMS}:
\eq{ \label{tra}
KI_{[2,2k+1]} = 
\mathop{\sum}_{\alpha,\beta,\gamma}(\Phi_c R^{2k+1} \Phi_c^{-1})^{\alpha\beta\beta}_{\alpha\gamma\gamma},
}
}
where $R$ is in our notation equal to
\eq{ \label{rm}
R = e^{\pi i \hbar\cdot A}
}
\NNN{
Now expanding the exponential in series, substituting formula \re{ddu1} for $\Phi_c$ and taking into account relations \re{relc}, one can arrive at the formula
\eq{
KI_{[2,2k+1]}= \kappa_1 + \kappa_2 A + \kappa_3 B + \kappa_4 AB +\kappa_5 BA + \kappa_6 ABA,
}
where $\kappa_1,\dots,\kappa_6$ are certain expressions in terms of the coefficients $\Phi_1,\dots,\Phi_6$ from formula \re{ddu1}.

Then, after taking traces of the individual terms in a way anologous to formulas (\ref{trA}-\ref{trBA}) and substituting $\Phi_1,\dots,\Phi_6$ from formula \re{relc}, we get}
\eq{
KI_{\NNN{[2,2k+1]}} = N\dfrac{\cos(\pi\hbar(2k+1))\sin(\pi\hbar)\cos(\pi\hbar N) + i\cdot
\sin(\pi\hbar(2k+1))\cos(\pi\hbar)\sin(\pi\hbar N)}{\cos(\pi\hbar)\sin(\pi\hbar N)} \label{tk}
}
Note that this answer is not normalized with respect to the unknot. Let us introduce
$q=e^{\pi i\hbar}$ and normalize with respect to the unknot:
\eq{
\label{pr1}
KI_{\NNN{[2,2k+1]}}/KI(\bigcirc) =
\dfrac{q^3\br{q^{2k+2}-q^{-2N+2k}+q^{-2N-2k}-q^{-2k-2}}}{q^4-1}
}
\NNN{
This expression is in a perfect agreement with known results. For example, it is known that in the Abelian case $N=1$ the expectation value of the Wilson 
loop is given by $q^{l(K)}$, where $l(K)$ is the so called  linking number of the knot $K$ \cite{Sm2}. The linking number can be computed as a sum of the orientations for the crossings of the knot, which takes values $\pm 1$.  In our case - Figure \ref{tref}, the knot has $2 k+1$ crossings with equal orientations, so the linking mumber is $l=2 k+1$. It is an elementary exercise to check that for $N=1$ our formula (\ref{pr1}) indeed gives $q^{2 k +1}$. In the first nontrivial case $N=2$ one can check that the result (\ref{pr1}) for any choice of $k$ is a Laurent polynomial in $q$ which coincide precisely with the Jones polynomial for the torus knot of type $[2,2k+1]$. For a general $N$ the answer (\ref{pr1}) is a polynomial in $A=q^{-N}$ and a Laurent polynomial in $q$. This two parametric polynomials again coincide with known HOMFLY polynomials for these knots as computed for example in \cite{kitaj,DBMMS}.  Being a rather nontrivial computation, this provides a strong test of our formulas for the associator.
}

\section{The Drinfeld prepotential}
In the previous sections we have computed the Drinfeld associator for the case of the fundamental representation explicitly and checked that it reproduces correct HOMFLY polynomials for toric the knots of type $[2,2k+1]$. In fact, as we explain below, the Drinfeld associator contains more information than is used for the knot invariants in the following sense: not all components of the Drinfeld associator contribute to the answer for the knot invariants.
 
 Using the perturbation approach to the solution of the regularized Knizhnik-Zamolodchikov equation (as in \ref{a:pert}), T.Le and J. Murakami found the expression for the associator as infinite sum in non-commutative operators $A$ and $B$ (without any additional relations of the form (\ref{rel}), i.e. for all representations) with the coefficients in the form of the multiple zeta function  values  \cite{Le}:
\be
\nonumber
\Phi_{3}=1^{\otimes 3} + \sum\limits_{k=2}^{\infty} \Big(\dfrac{\hbar}{2\pi i}\Big)^{k} \sum\limits_{m \geq 0} \sum\limits_{{\textbf{p}>0\,\textbf{q}>0 \atop | \textbf{p} |+| \textbf{q} |=k} \atop l( \textbf{p} )=l( \textbf{q} )=m } (-1)^{|\textbf{q}|} \, \tau(p_1,q_1,...p_m,q_m)\times\\
\sum\limits_{l(\textbf{r})=l(\textbf{s})=m \atop {0\leq \textbf{r} \leq \textbf{p},\, 0\leq \textbf{s}\leqq \textbf{q}}}\,(-1)^{|\textbf{r}|}\,\left( \prod\limits_{i=1}^{m} \Big( {{p_i}\atop{r_i}} \Big)\,\Big( {{q_i}\atop{s_i}} \Big) \right)\,B^{|\textbf{s}|}\,A^{p_1-r_1} B^{q_1-s_1}...A^{p_m-r_m} B^{q_m-s_m}A^{|\textbf{r}|}
\label{ass}
\ee
Where $\textbf{p}=(p_1,p_2,...,p_m)$ is a vector with positive integer components. The length of the vector $l(\textbf{p})=m$ and $|\textbf{p}|=\sum p_{i}$. For $\textbf{p}$ and $\textbf{q}$ the with positive integers $\textbf{p}>\textbf{q}$ means that $p_{i}>q_{i}$ and $\textbf{p}>0$ means that $p_{i}>0$ for all $i$. The coefficients $\tau(p_1,q_1,...p_m,q_m)$ are expressed through the multiple zeta functions as follows:
\be
\tau(p_1,q_1,...p_m,q_m)=\zeta(\underbrace{1,...,1}_{p_1-1},q_1+1,\underbrace{1,...,1}_{p_2-1},q_2+1,...,q_n+1)
\ee
such that for example $\tau(1,2)=\zeta(3)$ and $\tau(2,1)=\zeta(1,2)$. The zeta functions are defined as:
\be
\label{zeta1}
\zeta(m_1,m_2,...,m_n)=\sum\limits_{0<k_{1}<k_{2}<...<k_{n}}\,k_{1}^{-m_1}k_{2}^{-m_2}...k_{n}^{-m_n} \ee
The operators $A$ and $B$ depend on the gauge group and its representation:
\be
A=\Omega \otimes 1, \ \ \ B=1\otimes \Omega,\ \ \ \Omega=T_{R}^{a}\otimes T_{R}^{a}
\ee
\NNN{Formula (\ref{ass}) gives exact answer for the Drinfeld associator in all orders of $\hbar$, however it is impractical for explicit computations of the
knot invariants. For such computations one needs to summate exactly series (\ref{ass}) in the presence of some relations among  $A$ and $B$ given by a representation of the gauge group. In the case considered here (fundamental representation of $gl(N)$)  these relations are given by (\ref{rel}), (\ref{tb}) or \re{relc} depending on the associator type.   The main difficulty is that the values of the multiple zeta functions  (\ref{zeta1}) are difficult  to compute (see for example \cite{zud}) and there are no chances to summate series (\ref{ass}) exactly  in all orders of $\hbar$ by "bare hands" even in the presence of additional relations for $A$ and $B$.

Our approach implements the relations from the very beginning: using them, we reduce the KZ equation to the system of the hypergeometric equations. The solution of this system at the point $z=1$ gives the associator. In this way, we were able to summate the series (\ref{ass}) with nontrivial zeta functional coefficients explicitly and the result has the form of simple gamma functional factors (\ref{ans11}). Using computer we checked explicitly that up to the order $\hbar^5$ our result coincide precisely with series coming from (\ref{ass}). This should hold for all higher orders as well but we do not know how to show it explicitly due to the multiple zeta functions being difficult to work with, as mentioned above. }

\NNN{It is known, that the knot invariants must be rational functions in $q=\exp( i \pi \hbar)$ for any choice of gauge group and its representation. For example our answers for unknot \re{qdim} and torus knots of type $[2, 2k+1]$ (\ref{tk}) are obviously of this type. Therefore, the $\hbar$-series expansion of knot invariants has the form  $\sum_{k=0}^{\infty} a_{k}  h^k$ where we denoted $h=i \pi \hbar$, with \textit{rational} coefficients $a_{k}\in \mathbb{Q}$.}

The even values of the multiple zeta functions, i.e. $\zeta(m_1,m_2,...,m_n)$ for $k=m_1+m_2,...+m_n$-even number, are known to have the form $a \pi^k$ where $a$ is a rational number $a\in \mathbb{Q}$. On the other hand the odd zeta function values are always \textit{irrational} (conjecturally are transcendental  \cite{zud}.). \NNN{Moreover, the odd zeta values are algebraically independent over rational numbers, i.e. there is no polynomial relations among this values over $\mathbb{Q}$ ( the coefficient of polynomials are rational) \cite{zud}.}
   
   It is clear from the above formula (\ref{ass}) that the odd (even) zeta values represents the odd (even) coefficients of the Drinfeld associator expanded in the Planck's constant $\hbar$. Therefore, in order to make the coefficients of the $\hbar$-series for invariants \textit{rational}, the odd part of the associator must cancel in the answer. \NNN{Indeed, the knot invariant is given by a trace (over knot projection to two-dimensional plane) of a product of associators and brading matrices $R$ as for example in (\ref{tra}). Brading matrix is given by exponent (\ref{rm}) and its coefficients of $h$-expansion are rational for all representations.
Therefore, the contribution of the odd part of associator to the knot invariant is given by polynomial combinations of odd zeta values with rational coefficients, and as mentioned above such a combination is rational only if all the coefficients of the polynomial are zero, i.e. the coefficients of odd zeta values cancels.}

\NNN{Calculations up to $h^5$ shows the following:} every time we are computing the knot invariant by taking the trace of a braiding, for example as in (\ref{tra}), the odd zeta values enter the answer with a group factor (i.e. traces words made from $A$ and $B$) that are exactly zero due to celebrated STU and IHX relations for chord diagrams \cite{BRN}.
Therefore, for knot theory, we can use only the even part of the associator. We found, that dropping the odd terms in some clever way, the expression for associator possess valuable simplification. Indeed, consider the logarithm $F$ of the associator
\be
\Phi=\exp(F(\NNN{\hbar}))
\ee
Introduce then the symmetrized associator
(i.e. we drop all odd terms):
\be
\Phi_s=\exp\Big( \dfrac{1}{2}(F(\hbar)+F(-\hbar))\Big),\ \ \
\ee
\NNN{It should be clear from the discussion above}, that this "even" associator leads to the same knot invariants, but its structure is simpler. For example, in the fundamental representation case, for the associator of type (a) specified by the relations (\ref{rel}):
\eq{
A^2=B^2=\br{AB}^3=\br{BA}^3=1. 
}
we obtain:
\be
\Phi_{s}=\exp\Big({\cal{F}}_{s}(\hbar^2) [A,B]\Big)
\ee
with:
\be
{\cal{F}}_{s}(\hbar^2)=\dfrac{i}{2 \sqrt{N^2-1}}
\ln  \left( {\frac {i \left( {q}^{N}+1 \right)  \left( 1-q \right) N-2
\,N\sqrt { \left( {q}^{N+1}-1 \right)  \left( {q}^{N}-q \right) }}{
 \left( 1+q \right)  \left( 1-{q}^{N} \right)  \left( \sqrt {{N}^{2}-1
}+i \right) }} \right),\ \ \ q=e^{2 \pi i\hbar}
\ee
Therefore, the symmetrized associator is characterized by one function ${\cal{F}}_{s}$ that is even in $\hbar$, and has rational coefficients of $\hbar$ series expansion. We call this function Drinfeld prepotential, due to the role that the logarithm of the four point function plays in CFT, integrable systems and 4-dimensional SUSY models \cite{MS}.

\section{Conclusion}
To conclude, we solved the regularized Knizhnik-Zamolodchikov equation and found the explicit expressions for associators in the case of the fundamental representation of $gl(N)$. It turned out that associator is a finite series in $A$ and $B$ with coefficients given by values of hypergeometric functions taken at the point $z=1$. Also we presented some tests of our results.

One possible direction of further research is to consider higher representations. For that case everything becomes more complicated due to more complicated relations for $A$ and $B$.

It would also be very interesting to understand the relation between conformal blocks and associator more deeply.

\section*{Acknowledgements}
We are grateful to A.Mironov and Ye.Zenkevich for helpful remarks. We are especially grateful to A.Morozov for stimulating discussions.

We are also very grateful to the referees of Journal of Physics A: Mathematical and Theoretical for pointing out several flaws in the original version of the paper.

Our work is partly supported by Ministry of Education and Science of the Russian Federation under contracts
14.740.11.0081 (P.DB., A.Sl.) and 14.740.11.0347 (A.Sm.), by RFBR grants 10-02-00499 (P.DB.) and 10-01-00536 (A.Sl., A.Sm.), by joint grants 12-02-92108-YaF ( A.Sl.), 12-02-91000-ANF (A.Sl.), 11-01-92612-Royal Society(P.DB., A.Sl., A.Sm.), by NWO grant 613.001.021 (P.DB.), by RFBR-09-01-93106-NCNILa (A.Sm.) and by Dynasty foundation.

\appendix
\section{Solution of KZ equation in the first two orders of perturbative expansion}
\label{a:pert}
In this appendix we solve KZ equation in the first two orders of perturbative expansion. For KZ equation
$$
\dfrac{d G(z)}{d z} =\hbar\,\Big( \dfrac{A}{z}+\dfrac{B}{1-z} \Big)\,G(z),
$$
let us try to find the solution as a series in $\hbar$:
$$
G(z)=G_0(z)+\hbar\, G_{1} (z)+\hbar^2\, G_{2} (z)+\dots
$$
Equation (\ref{KZ}) has a unique solution defined by boundary condition:
$$
G_0(0)=1=1\otimes1\otimes1
$$
We are interested in the value $G(1)$.

First order:
$$
\dfrac{d G_1(z)}{d z}=\dfrac{A}{z}+\dfrac{B}{1-z}\ \ \Rightarrow \ \ G_1 (1)=A
\int\limits_{0}^{1} \dfrac{d z}{z} +B \int\limits_{0}^{1} \dfrac{d z}{1-z}
$$
The integrals here are divergent so we normalize them by:
$$
\int\limits_{0}^{1}\rightarrow \int\limits_{\epsilon}^{1-\epsilon}
$$
Then we have:
$$
G_{1}(z)=A \log\Big(\dfrac{z}{\epsilon}\Big) - B \log
\Big(\dfrac{1-z}{1-\epsilon}\Big)
$$
and, keeping only finite and singular terms at point $z=1$,
$$
G_{1}(1)= - A \log(\epsilon) - B \log (\epsilon)
$$

Second order:
\eq{
\dfrac{d G_2 (z)}{d z}=\br{\dfrac{A}{z}+\dfrac{B}{1-z}} G_1(z)=\\
=A^2 \dfrac{1}{z} \log\dfrac{z}{\epsilon}-AB \dfrac{1}{z} \log\dfrac{1-z}{1-\epsilon}+BA \dfrac{1}{1-z} \log\dfrac{z}{\epsilon}-B^2 \dfrac{1}{z} \log\dfrac{1-z}{1-\epsilon}
}
Integrating over $[\epsilon, z]$, we get:
\eqm{
G_2(z) = \dfrac{1}{2} A^2 \log^2\dfrac{z}{\epsilon}-AB\br{\Li_2(\epsilon)-\Li_2(z)-\log(1-\epsilon)\log\dfrac{z}{\epsilon}} +\\
+BA\br{\Li_2(\epsilon)-\Li_2(z)-\log(1-z)\log\dfrac{z}{\epsilon}}+\dfrac{1}{2} B^2 \log^2\dfrac{1-z}{1-\epsilon}
}
At point $z=1-\epsilon$, for finite and singular terms we have:
\eq{
G_2(1) = \dfrac{1}{2} A^2 \log^2\epsilon + BA \log\epsilon\log\epsilon + \dfrac{1}{2} B^2 \log^2\epsilon + [A,B] \Li_2(1)
}
Thus, up to the second order in $\hbar$, $G(1)$ has the following form (we drop the terms which tend to zero as $\epsilon \rightarrow 0$):
\eq{
G(z) = \br{1-\hbar B \log\epsilon+\dfrac{1}{2}\hbar^2 B^2 \log^2\epsilon}\br{1+\hbar^2 [A,B] \Li_2(1)}\br{1-\hbar A \log\epsilon+\dfrac{1}{2}\hbar^2 A^2 \log^2\epsilon} + \Ob\br{\hbar^3}
}

\end{document}